\documentclass[aps,prl,reprint,preprintnumbers]{revtex4-2}

\usepackage[english]{babel}

\usepackage{amsmath}
\usepackage{graphicx}
\usepackage{tikz}
\usepackage{subfigure}
\usepackage[colorlinks=true, allcolors=blue]{hyperref}

\usepackage{amssymb}
\usetikzlibrary{shapes.geometric,arrows,arrows.meta,decorations.pathmorphing,decorations.markings,patterns}

\usetikzlibrary{decorations.markings}
\usetikzlibrary{decorations.pathmorphing}
\usetikzlibrary{intersections}
\usetikzlibrary{decorations.text}
\usetikzlibrary{calc}

\newcommand{\ep}{\epsilon}
\newcommand \vev [1] {\langle{#1}\rangle}
\begin{document}
\preprint{LAPTH-035/25, MPP-2025-196, USTC-ICTS/PCFT-25-48}
\title{Bootstrapping Six-Gluon QCD Amplitudes}

\author{Sérgio Carrôlo$^{a}$}
\email{scarrolo@mpp.mpg.de}
\author{Dmitry Chicherin$^{b}$}
\email{chicherin@lapth.cnrs.fr}
\author{Johannes Henn$^{a}$}
\email{Corresponding author. E-Mail: henn@mpp.mpg.de.}
\author{Qinglin Yang$^{a}$}
\email{qlyang@mpp.mpg.de}
\author{Yang Zhang$^{c,d,e}$}
\email{yzhphy@ustc.edu.cn}

\affiliation{
$a$ Max-Planck-Institut f\"{u}r Physik, Werner-Heisenberg-Institut, Boltzmannstr. 8,
85748 Garching, Germany \\
$b$ LAPTh, Universit\'e Savoie Mont Blanc, CNRS, B.P. 110, F-74941 Annecy-le-Vieux, France\\
$c$ Interdisciplinary Center for Theoretical Study, University of Science and Technology of China, Hefei, Anhui 230026, China\\
$d$ Peng Huanwu Center for Fundamental Theory, Hefei, Anhui 230026, China\\
$e$ Center for High Energy Physics, Peking University, Beijing 100871, People’s Republic of China
}

\begin{abstract}
We present a symbol-level bootstrap construction of the planar, two-loop six-gluon scattering amplitude for the $\mathord{-}\mathord{-}\mathord{+}\mathord{+}\mathord{+}\mathord{+}$ helicity configuration in Quantum Chromodynamics (QCD), focusing on the maximal weight pieces—the ``most complicated terms'' in the sense of Lipatov et al. 
Building on recent advances in the understanding of the relevant function space, we incorporate as a crucial new ingredient the complete set of leading singularities, obtained from an 
analysis of on-shell diagrams. The resulting expressions are manifestly conformally invariant and clarify the structure of previous five-particle results. 
Combining this with the symbol bootstrap, 
we show that constraints from physical limits are sufficient to  
determine the answer. 
We thus obtain the first concrete characterization of two-loop six-gluon amplitudes at symbol level and at highest weight. 
Remarkably, we find that the effective function space involves only 137 symbol letters—significantly fewer than the full set of 
167 
possible letters, suggesting a yet-unexplained underlying structure akin to that seen in maximally supersymmetric Yang–Mills theory. 
From the novel amplitude results we extract  previously unknown symbol-level results describing two-loop triple collinear and double soft limits.
\end{abstract}

\maketitle

\section{Introduction}

%Scattering amplitudes are fundamental objects that describe how particles interact in quantum field theory. Beyond their practical role in the phenomenology of high-energy physics, they have revealed hidden structures of gauge theories, with deep connections to mathematics. These insights have also led to radically novel conceptual approaches, with important consequences for our ability to compute amplitudes. Examples include generalized unitarity, on-shell methods, the Grassmannian formulation of leading singularities, as well as advances in understanding the special functions arising from the Feynman integrals. For recent reviews, cf. \cite{Arkani-Hamed:2012zlh,Elvang:2015rqa,Badger:2023eqz}.

Scattering amplitudes are fundamental objects describing particle interactions in quantum field theory. Beyond their role in high-energy phenomenology, they have revealed hidden structures of gauge theories and deep connections to mathematics, leading to radically new conceptual approaches to amplitude computation. Notable examples include generalized unitarity, on-shell methods, the Grassmannian formulation of leading singularities, and advances in understanding the special functions arising from Feynman integrals. For recent reviews, see refs.~\cite{Arkani-Hamed:2012zlh,Elvang:2015rqa,Badger:2023eqz}.

%The amplitude bootstrap \cite{Bern:1994cg,Caron-Huot:2020bkp} successfully combines ideas from the S-matrix program with these modern advances. In maximally supersymmetric Yang–Mills theory (sYM), bootstrap methods based on the symbol of iterated integrals \cite{Brown:2009qja,Goncharov:2010jf} have enabled the determination of six-gluon scattering amplitudes through high loop orders. This progress has led to a detailed understanding of their analytic behavior in various kinematic limits, such as the collinear and Regge regimes, and uncovered striking connections to integrability, cluster algebras, and leading singularities. 

The amplitude bootstrap \cite{Bern:1994cg,Caron-Huot:2020bkp} combines modern 
\mbox{S-matrix} ideas with these developments. In maximally supersymmetric Yang–Mills theory (sYM), bootstrap methods based on the symbol of iterated integrals \cite{Brown:2009qja,Goncharov:2010jf} have enabled the determination of six-gluon scattering amplitudes through high loop orders, leading to a detailed understanding of their analytic structure in kinematic limits such as collinear and Regge regimes, and uncovering connections to integrability, cluster algebras, and leading singularities.

%It is an open problem to extend this program to QCD scattering amplitudes beyond the one-loop order. In this work we take a first step in this direction: we apply the symbol bootstrap, for the first time, to a six-particle scattering amplitude in massless QCD.
%In this way we provide first results beyond the current state-of-the-art, which is at two loops and five particles \cite{Huss:2025nlt}. 

Extending bootstrap methods to QCD scattering amplitudes beyond one loop remains an open challenge. In this Letter, we take a first step in this direction by applying the symbol bootstrap, for the first time, to a six-particle scattering amplitude in massless QCD. This goes beyond the current state of the art, which is limited to two loops and five particles \cite{Huss:2025nlt}. 
%As a proof of principle, we focus on the leading-color two-loop 
%$\mathord{-}\mathord{-}\mathord{+}\mathord{+}\mathord{+}\mathord{+}$ helicity amplitude, parametrized as
We parametrize the amplitude as
\begin{align}
{\cal A} = \sum_{ij} c_{ij} R_i f_j \,,
\label{eq:ansatz}
\end{align}
%The $L$-loop correction to a scattering amplitude in perturbative QCD can be schematically parametrized as
%\begin{align}\label{eq:ansatz}
%{\cal A} = \sum_{ij} c_{ij} R_{i} f_{j} \,.
%\end{align}
%Both the set of special functions $f$ and the prefactors $R$ depend on the scattering kinematics.
%In the (symbol) bootstrap, the $c$ are a set of coefficients that are to be determined.
%In this Letter, as a proof of principle, we focus on the leading-color two-loop $\mathord{-}\mathord{-}\mathord{+}\mathord{+}\mathord{+}\mathord{+}$ helicity amplitude.
%The relevant set of special functions $f$ is known from refs. \cite{Abreu:2024fei,Henn:2025xrc}. 
%They can be written as iterated integrals that depend on seven kinematic variables associated to the six-particle scattering, and involve up to $167$ symbol letters. At two loops, iterated integrals (or symbols) of up to length (or weight) four are expected to appear. The challenge is to determine within this large set of functions the relevant one that describes the scattering amplitude. 
where the special functions $f_j$ and rational prefactors $R_{i}$ depend on the scattering kinematics, while $c_{ij}$ are the ansatz coefficients.
%determined by imposing physical constraints. 
The relevant %six-particle 
function space is known up to two loops \cite{Abreu:2024fei,Henn:2025xrc} in the planar case, consisting of iterated integrals in seven kinematic variables with up to 167 symbol letters and maximal weight four. However, this knowledge alone does not determine the amplitude: the central challenge is to identify the specific linear combination selected by QCD.
%
%Given the tremendous advances in our ability of computing Feynman integrals, a new bottleneck concerns the prefactors $R_{i}$. Even writing down the known results for five-particle prefactors in a concise way is challenging, see e.g. \cite{Heller:2021qkz,DeLaurentis:2025dxw}.
%Remarkably, in maximally sYM, the latter can be obtained from four-dimensional leading singularities \cite{Arkani-Hamed:2012zlh}. Predicting the allowed set of prefactors in QCD by independent methods is an important open problem.

A major bottleneck in this program concerns the rational prefactors 
$R_{i}$. Even at five points, their explicit form is highly non-trivial \cite{Heller:2021qkz,DeLaurentis:2025dxw}. In sYM, these prefactors are governed by four-dimensional leading singularities \cite{Arkani-Hamed:2012zlh}, but no comparable organizing principle has been established for QCD.

%In this Letter, we focus on the ``most complicated terms'' in eq. (\ref{eq:ansatz}), in the sense of Lipatov {\it{et al}} \cite{Kotikov:2002ab,Kotikov:2004er}.
%These authors noticed that in the context of Regge physics the highest-weight polylogarithmic terms agree between QCD and sYM. 
%This ``maximal transcendentality principle'' also extends to certain anomalous dimensions \cite{Dixon:2017nat} and form factors \cite{Guo:2022pdw},
%but not to generic scattering amplitudes \cite{Gehrmann:2011xn}.
%We show that, while the maximally transcendental  terms of QCD and sYM amplitudes differ, the prefactors multiplying the highest-weight functions in QCD are particularly simple and can be accessed using tools developed in sYM.
%We reveal that they are fully captured by four-dimensional leading singularities, possess a conformal symmetry, and are given by concise spinor-helicity expressions.

In this Letter, we address this problem by focusing on the maximally transcendental contributions to eq.~\eqref{eq:ansatz}, in the sense of Lipatov {\it et al.}~\cite{Kotikov:2002ab,Kotikov:2004er}. Although the maximal-weight parts of QCD and sYM scattering amplitudes do not generally coincide \cite{Gehrmann:2011xn}, we show that the prefactors multiplying the highest-weight functions in QCD are remarkably simple. They are entirely captured by four-dimensional leading singularities, exhibit a conformal symmetry, and admit compact spinor-helicity representations. A relation between QCD prefactors and four-dimensional on-shell structures was previously observed 
%in a different context 
in ref.~\cite{Henn:2021aco}, but its broader implications for bootstrapping  multi-particle amplitudes were not explored.

%Our results establish a new link between QCD amplitudes and on-shell methods developed in maximally supersymmetric Yang–Mills theory. We begin by identifying the relevant four-dimensional leading singularities in pure Yang–Mills theory, revealing their remarkable simplicity and conformal invariance. Motivated by this structure, we conjecture that these leading singularities form a complete basis for the rational prefactors multiplying maximally transcendental functions in QCD. Imposing physical constraints within this bootstrap framework, we then uniquely determine the weight-four symbol of the leading-color two-loop 
%$\mathord{-}\mathord{-}\mathord{+}\mathord{+}\mathord{+}\mathord{+}$ planar six-gluon amplitude. This represents the first successful application of the symbol bootstrap to gluon scattering amplitudes in massless QCD beyond one loop.

Building on this structure, we conjecture that these leading singularities form a complete basis for the rational prefactors at maximal transcendental weight. 
By imposing physical constraints within this bootstrap framework, we uniquely determine the weight-four symbol of the leading-color two-loop
$\mathord{-}\mathord{-}\mathord{+}\mathord{+}\mathord{+}\mathord{+}$ planar six-gluon amplitude. This constitutes the first successful application of the symbol bootstrap to gluon scattering amplitudes in massless QCD beyond one loop. 
As a byproduct, we extract the symbols of previously unknown triple-collinear splitting and double-soft functions.
A companion paper provides more details and extends this proof of principle to all other MHV helicity configurations \cite{Carrolo:2026qpu}.

Our maximal-weight strategy is conceptually straightforward to extend to arbitrary multiplicity and to non-planar contributions, although its practical implementation becomes more computationally demanding and is not pursued here. The extension to lower-weight terms is more subtle, and we comment on this issue and outline concrete avenues for future investigation in the outlook. More broadly, when combined with Landau analysis to constrain the relevant singularity structure, our approach has the potential to bypass the explicit computation of Feynman integrals altogether.
% While our analysis focuses on the maximally transcendental sector, the leading-weight strategy generalizes straightforwardly to arbitrary multiplicity and to non-planar contributions, although we do not pursue these directions here. The extension to lower-weight terms is more subtle; we comment on this issue and outline concrete avenues for future investigation in the outlook. As a byproduct, we extract the symbols of previously unknown triple-collinear splitting and double-soft functions. More broadly, when combined with Landau analysis to constrain the relevant singularity structure, our approach has the potential to bypass the explicit computation of Feynman integrals altogether.

%We begin by discussing the necessary leading singularities in pure Yang--Mills theory (YM), revealing their simplicity and conformal invariance. By making a conjecture that they form a complete basis for all prefactors at maximal transcendental weight, we then present the ingredients and physical conditions that enter the bootstrap approach, and use these to completely fix the weight-four symbol of the $\mathord{-}\mathord{-}\mathord{+}\mathord{+}\mathord{+}\mathord{+}$ planar two-loop scattering amplitude. This is the first time the bootstrap approach has been successfully applied to gluon scattering amplitudes in massless QCD beyond one loop. 
%As a byproduct, we extract the corresponding symbols of the previously unknown triple collinear splitting and double soft functions. We close with a discussion of new opportunities opened up by this research.

\section{Leading singularities and prefactors at maximal weight}

In this section, we discuss the prefactors multiplying maximal-weight terms in scattering amplitudes, following ref.~\cite{Henn:2021aco}.
Generalized unitarity suggests that such prefactors are governed by leading singularities.
An important subtlety arises, however, in dimensionally regulated amplitudes: even at maximal weight, additional kinematic prefactors can appear beyond the four-dimensional leading singularities. This phenomenon was observed in sYM and supergravity amplitudes at two loops, \cite{Abreu:2018aqd,Chicherin:2018old,Chicherin:2019xeg,Abreu:2019rpt}. There,
the ${\cal{O}}(\epsilon^0)$ term at two loops contains additional prefactors, which originate from leading singularities evaluated in $4{-}2\epsilon$ dimensions.
However, those prefactors cancel for appropriate infrared-subtracted hard functions. (The definition will be given in the next section.) We will therefore adopt the conjecture that the four-dimensional leading singularities form a complete basis for prefactors in our ansatz for the hard function.

The leading singularities can be computed from on-shell diagrams built from three-gluon on-shell vertices \cite{Arkani-Hamed:2012zlh}. 
It can be shown that this on-shell ``gluing'' procedure preserves conformal invariance. This is in line with the classical conformal invariance of massless QCD theory.
Therefore, we are guaranteed to obtain formulas that are annihilated by the operator \cite{Witten:2003nn},
\begin{align}\label{eq:Wittenconformaloperater}
k_{\alpha \dot \alpha} = \sum_{i=1}^{n} \frac{\partial}{\partial \lambda_{i}^{\alpha}} \frac{\partial}{\partial \tilde\lambda_{i}^{\dot \alpha}} \,,
\end{align}
where we use the usual spinor-helicty parametrization of momenta, $p_i^{\alpha \dot\alpha} = \lambda^{\alpha}_i \tilde{\lambda}_i^{\dot\alpha}$, and $\vev{ij} = \lambda_i^{\alpha} \lambda_{j\,\alpha}$.
This conformal invariance is trivially seen for the tree-level amplitude, which is given by the following Parke-Taylor factor,
\begin{align}\label{eq:PTmmpppp}
{\rm PT}_{1,2} = \frac{\vev{12}^4}{\vev{12}\vev{23} \vev{34} \vev{45} \vev{56}\vev{61}} \,,
\end{align}
and it also holds for all known expressions for one-loop six-gluon and six-photon amplitudes \cite{DHMZ2021}.

Let us focus on the $\mathord{-}\mathord{-}\mathord{+}\mathord{+}\mathord{+}\mathord{+}$ six-gluon case from now on. Our goal is to predict the relevant leading singularities at two loops.

The one-loop $\mathord{-}\mathord{-}\mathord{+}\mathord{+}\mathord{+}\mathord{+}$ helicity amplitudes were first calculated in ref. \cite{Bern:1994cg}. We can see from the explicit formulas that the coefficients of the highest weight terms are simply given by eq. (\ref{eq:PTmmpppp}). Indeed, this is consistent with evaluating the relevant on-shell box diagrams.

\begin{figure}[t!]
    \centering
    \includegraphics[scale=0.6]{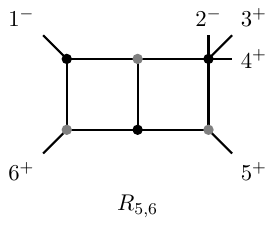}
    \includegraphics[scale=0.6]{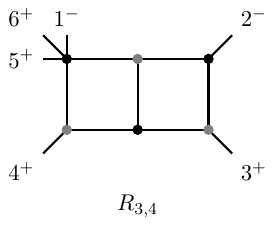}
    \includegraphics[scale=0.6]{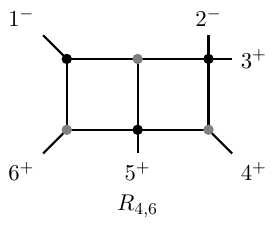}
    \includegraphics[scale=0.6]{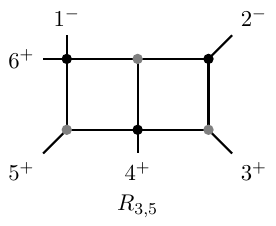}
    \includegraphics[scale=0.6]{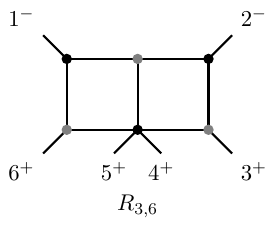}
    \includegraphics[scale=0.6]{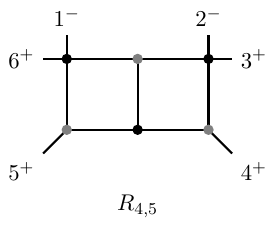}
    \caption{On-shell diagrams giving rise to leading singularities $R_{i,j}$,  
    for the $\mathord{-}\mathord{-}\mathord{+}\mathord{+}\mathord{+}\mathord{+}$ helicity configuration.
    }
    \label{fig:LSpart1}
    \vspace{-3ex}
\end{figure}

At two loops, the Parke-Taylor prefactor $R_1 ={\rm PT}_{1,2}$ is supplemented by six new leading singularities. These can be calculated from two-loop generalized unitarity, cf. Fig.~\ref{fig:LSpart1} for the relevant on-shell diagrams.
We evaluate these using techniques explained in ref. \cite{Arkani-Hamed:2012zlh}. For instance, 
the two-loop diagram for $R_{5,6}$ evaluates to the following one-form,
\begin{equation}  \label{eq:oneform} \text{PT}_{1,2}\times\frac{z^3(\langle12\rangle{+}z\langle26\rangle)^3\langle56\rangle^4}{\langle12\rangle^4(\langle15\rangle{+}z \langle56\rangle)^4} \, dz \,.
\end{equation}
The non-trivial two-loop leading singularity $R_{5,6}$ is then obtained by computing the value at one of the residues of eq. (\ref{eq:oneform}).
It turns out that same formula applies to all the leading singularities obtained from  Fig.~\ref{fig:LSpart1}, and that the result can be written as follows,
\begin{align}\label{eq:twoloopLScompact}
{R_{i,j}}/{R_{1}} =& {-}1 + 
 12  u_{i,j}  -  30 u_{i,j}^2 + 20 u_{i,j}^3 
  \,,
 \end{align}
with $u_{i,j}:= \vev{ 1  i } \vev{2  j} /( \vev{1  2}  \vev{ij})$ 
Note that eq. (\ref{eq:twoloopLScompact}) is expressed in terms of $\lambda$ spinors only, and hence conformal invariance is manifest. 

We remark that eq. (\ref{eq:twoloopLScompact}) with $2<i<j\leq n$ also captures all two-loop leading singularities with $--+\cdots+$ helicity configurations for arbitrary $n$. 
As a consistency check, one may verify that the known maximal-weight coefficients in the five-particle two-loop amplitudes, cf. refs. \cite{Abreu:2019odu,Agarwal:2023suw}, can be simplified to the expressions given here.

\section{Amplitude Symbol Bootstrap}

We now compute the symbol of the leading-color two-loop $\mathord{-}\mathord{-}\mathord{+}\mathord{+}\mathord{+}\mathord{+}$ six-gluon amplitude in pure YM theory. 
We consider an infrared-finite hard function \cite{Catani:1998bh},
\begin{align}
{\cal H}^{(2)} = 
\underset{\ep \to 0}{{\rm lim}}\left[
{\cal A}^{(2)} - I^{(1)} {\cal A}^{(1)} -
I^{(2)} {\cal A}^{(0)} \right]
\,.
\label{eq:Ampl_IR_subtrac}
\end{align}
Here  ${\cal A}^{(L)}(\ep)$ are planar color-stripped $L$-loop amplitudes.
$I^{(1)}(\ep)$ is given by \footnote{Note that switching e.g. to minimal subtraction, $I^{(1)}_{min,n}:=-\frac{n}{\epsilon^2}{+}\frac{1}\epsilon\sum_{i=1}^{n}\log\left(\frac{-s_{i,i{+}1}}{\mu^2}\right)$, only results in finite differences to the hard function.},
\begin{equation}\label{eq:sub}
    I_n^{(1)}=-\frac1{\epsilon^2}\sum_{i=1}^n\left(\frac{-s_{i,i{+}1}}{\mu^2}\right)^{-\epsilon}\,,
\end{equation}
and, keeping only terms  relevant at symbol level and at highest weight,
$I^{(2)} = - \frac12 (I^{(1)})^2$, cf.  \cite{Aybat:2006mz}.
Ultraviolet divergences are known to affect lower weight terms only (at least, up to the loop order relevant here),
hence they do not influence the highest-weight approximation.

Our strategy to find the hard function ${\cal H}^{(2)}$ 
involves: 
\begin{enumerate}
    \item Choosing an appropriate ansatz;
    \item Imposing discrete symmetries on the ansatz;
    \item Requiring the physical singularity structure and consistency with the (multi)-collinear limits. 
\end{enumerate}
Remarkably, we find that the bootstrap assumptions, together with these physical conditions, not only fix the ansatz uniquely, but also, as a bonus, determine the previously unknown triple collinear splitting and double soft functions. 

Let us discuss the bootstrap procedure in more detail. 

1. Our bootstrap ansatz for the weight-four symbol terms of the two-loop hard function takes into account the seven leading singularities determined above,
\begin{align}\label{eq:ansatzhard}
{\cal H}_{\text{YM}}^{(2)} = R_1G_1{+} \sum_{2<i<j\leq6} R_{i,j} G_{i,j}   \,. 
\end{align}
The ans\"{a}tze for the $G_1$, $G_{i,j}$ are weight-four symbols of the two-loop six-leg Feynman integrals closed under cyclic permutations.
In order to have the relevant space of functions $G$, it is important to take into account eq. (\ref{eq:Ampl_IR_subtrac}).
 Since the tree-level and one-loop amplitudes involve only the $R_1$ prefactor, it follows that only $G_1$ is affected by the infrared subtractions. We take this into account by adding appropriate product terms of $I^{(L)}$ and lower-loop amplitudes to our ansatz. 

2.   
The flip symmetry $(123456) \leftrightarrow (216543)$
of the $\mathord{-}\mathord{-}\mathord{+}\mathord{+}\mathord{+}\mathord{+}$ helicity configuration 
reduces the ansatz size for $G$. 
Requiring the functions $G_1$ and $G_{i,j}$ to have the same transformation law as $R_1$ and $R_{i,j}$ leads to an ansatz \eqref{eq:ansatzhard} that contains $2412$ unknowns.

3. 
The rational prefactors $R_{i,j}$ 
contain spurious poles $1/\vev{ij}^3$ for nonadjacent $i,j$, and higher-order poles $1/\vev{ij}^4$ for adjacent $i,j$, which have to be suppressed by the vanishing of the symbol $G_{i,j}$ at the locus $\vev{ij} = 0$. We also take into account that symbol entries of $G$ have to be dimensionless.

If a pair of adjacent momenta become collinear, e.g. $p_5^{h_5} \to z\, p^h$ and $p_6^{h_6} \to (1{-}z)\,  p^h$,  an amplitude may become singular, in which case its leading terms factorize  \cite{Kosower:1999rx,Bern:1994zx,Bern:2004cz}. This can be schematically represented as follows, 
\begin{equation}
    \vcenter{\hbox{
 	\begin{tikzpicture}[line width=0.6 ,scale=0.75,line cap=round,every node/.style={font=\footnotesize}]
 			\draw[very thick] (0,0) circle(0.4);
 			\node (tt) at (0,0) {$\mathcal{A}_6$};
 		\begin{scope}[rotate=30]
 			\coordinate (p1) at ($(0.4,0)+1*(0.5,0)$);
 			\coordinate (p2) at ($(0.4,0)$);
  			\node (tt) at ($(p1)+(0.3,0)$) {$p_6$};
 			\draw[very thick] (p1) to (p2);
 		\end{scope}
 		\begin{scope}
 			\coordinate (p1) at ($(0.4,0)+1*(0.5,0)$);
 			\coordinate (p2) at ($(0.4,0)$);
  			\node (tt) at ($(p1)+(0.3,0)$) {$p_5$};
 			\draw[very thick] (p1) to (p2);
 		\end{scope}
 		\begin{scope}[rotate=-30]
 			\coordinate (p1) at ($(0.4,0)+1*(0.5,0)$);
 			\coordinate (p2) at ($(0.4,0)$);
 			 \node (tt) at ($(p1)+(0.3,0)$) {$p_4$};
			\draw[very thick] (p1) to (p2);
 		\end{scope}
 		\begin{scope}[rotate=150]
 			\coordinate (p1) at ($(0.4,0)+1*(0.5,0)$);
 			\coordinate (p2) at ($(0.4,0)$);
 			 \node (tt) at ($(p1)+(0.3,0)$) {$p_1$};
 			\draw[very thick] (p1) to (p2);
 		\end{scope}
 		\begin{scope}[rotate=180]
 			\coordinate (p1) at ($(0.4,0)+1*(0.5,0)$);
 			\coordinate (p2) at ($(0.4,0)$);
 			 \node (tt) at ($(p1)+(0.3,0)$) {$p_2$};
 			\draw[very thick] (p1) to (p2);
 		\end{scope}
 		\begin{scope}[rotate=210]
 			\coordinate (p1) at ($(0.4,0)+1*(0.5,0)$);
 			\coordinate (p2) at ($(0.4,0)$);
 			\node (tt) at ($(p1)+(0.3,0)$) {$p_3$};
 			\draw[very thick] (p1) to (p2);
 		\end{scope}
 	\end{tikzpicture}}}
 	\,
 	\begin{tikzpicture}[scale=0.6,every node/.style={font=\footnotesize}]
 		\begin{scope}
 				\draw[->, thick] (0,0) to (1,0)
 				node[above] at(0.5,0.1) {$p_6||p_5$};
 		\end{scope}
 	\end{tikzpicture}\  \,
     \hspace{-7pt}
 	\left(
        \hspace{-4pt}
 		\vcenter{\hbox{
 		\begin{tikzpicture}[scale=0.7,every node/.style={font=\footnotesize}]
 			\begin{scope}
 			\draw[very thick] (0,0) circle(0.4);
 			\node (tt) at (0,0) {$\mathcal{A}_5$};
 		\begin{scope}[rotate=30]
 			\coordinate (p1) at ($(0.4,0)+1*(0.5,0)$);
 			\coordinate (p2) at ($(0.4,0)$);
  			\node (tt) at ($(p1)+(0.3,0)$) {$p$};
 			\draw[very thick] (p1) to (p2);
 		\end{scope}
 		\begin{scope}[rotate=-30]
 			\coordinate (p1) at ($(0.4,0)+1*(0.5,0)$);
 			\coordinate (p2) at ($(0.4,0)$);
 			 \node (tt) at ($(p1)+(0.3,0)$) {$p_4$};
			\draw[very thick] (p1) to (p2);
 		\end{scope}
 		\begin{scope}[rotate=150]
 			\coordinate (p1) at ($(0.4,0)+1*(0.5,0)$);
 			\coordinate (p2) at ($(0.4,0)$);
 			 \node (tt) at ($(p1)+(0.3,0)$) {$p_1$};
 			\draw[very thick] (p1) to (p2);
 		\end{scope}
 		\begin{scope}[rotate=180]
 			\coordinate (p1) at ($(0.4,0)+1*(0.5,0)$);
 			\coordinate (p2) at ($(0.4,0)$);
 			 \node (tt) at ($(p1)+(0.3,0)$) {$p_2$};
 			\draw[very thick] (p1) to (p2);
 		\end{scope}
 		\begin{scope}[rotate=210]
 			\coordinate (p1) at ($(0.4,0)+1*(0.5,0)$);
 			\coordinate (p2) at ($(0.4,0)$);
 			\node (tt) at ($(p1)+(0.3,0)$) {$p_3$};
 			\draw[very thick] (p1) to (p2);
 		\end{scope}
 		\end{scope}
 	\end{tikzpicture}}}
    \hspace{-3pt}
 	\right)
     \hspace{-3pt}
 	 \times 
     \hspace{-3pt}
 	\vcenter{\hbox{\begin{tikzpicture} [scale=0.75,every node/.style={font=\footnotesize}]		
 		\begin{scope}
 			\coordinate (p1) at (-0.7,0);
 			\coordinate (p2) at ($0.7*({cos(60)},{sin(60)})$);
 			\coordinate (p3) at ($0.7*({cos(-60)},{sin(-60)})$);
 			\node (t1) at ($(p1)-(0.2,0)$) {$p$};
 			\node (t2) at ($(p2)+0.2*({cos(60)},{sin(60)})$) {$p_5$};
 			\node (t3) at ($(p3)+0.2*({cos(-60)},{sin(-60)})$) {$p_6$};

 			\draw[very thick] (0,0) circle (0.3);
 			\draw[very thick] (p1) to ($(-0.3,0)$);
 			\draw[very thick] (p2) to ($0.3*({cos(60)},{sin(60)})$);
			\draw[very thick] (p3) to ($0.3*({cos(-60)},{sin(-60)})$);
			\node (t) at (0,0) {$Sp$};
 		\end{scope}
 	\end{tikzpicture}}}
    \, .
 \end{equation}
For the two-loop hard function ${\mathcal H}^{(2)}_6 \equiv {\cal H}_{\text{YM}}^{(2)}$, this turns into the following equation at maximal transcendental weight level, 
\begin{equation}
    \begin{aligned}
 {\mathcal H}^{(2)}_6 /{\cal A}^{(0)}_6 \xrightarrow{p_5 || p_6}{} 
& %
 {\mathcal H}^{(2)}_{5}/ {\cal A}^{(0)}_{5} \\
& {+}\mathcal{C}_{h_5h_6}^{h,(1)} \,
{\mathcal H}^{(1)}_{5}/{\cal A}^{(0)}_{5}{+}\frac{1}{2} \left( \mathcal{C}_{h_5h_6}^{h,(1)} \right)^2,\label{eq:HAcoll}
\end{aligned}
\end{equation}
where ${\mathcal A}^{(0)}_n$ are tree-level color-stripped amplitudes, ${\mathcal H}^{(L)}_5$ are five-particle $L$-loop MHV hard functions, and $\mathcal{C}_{h_5h_6}^{h,(L)}$ are infrared-subtracted $L$-loop corrections to the splitting functions for helicity $h_5h_6\to h$. 
The $\mathord{+}\mathord{+}\to \mathord{-}$ splitting functions vanish at tree level, and their loop functions are of subleading transcendental weight, so we can neglect them in our study. The remaining splitting functions are (regardless of helicity)
\begin{equation}
    \begin{aligned}
\mathcal{C}_{h_5h_6}^{h,(1)} = & \log(1{-}z) \log\left((p_1 \cdot p)/(p_5 \cdot p_6)\right)
\\ & \hspace{-1.2cm} + \log(z) \log(1{-}z) + \log(z) \log\left( (p_4 \cdot p)/(p_5 \cdot p_6) \right) 
  \,.
\label{eq:C1}
\end{aligned}
\end{equation}
Likewise, for ${\mathcal H}^{(L)}_5$, calculated in refs. \cite{Abreu:2019odu,Agarwal:2023suw}, we need the $\mathord{-}\mathord{-}\mathord{+}\mathord{+}\mathord{+}$ helicity configuration only.  The number of constraints obtained from collinear limits for adjacent pairs of momenta (or, equivalently, taking $p_5||p_6$ for cyclic permutations of ${\cal H}_6$) are shown %in the second column of 
Table~\ref{tab:bootstrap}. 

\begin{table}[]
\caption{Two-loop bootstrap constraints for ${\cal H}^{(2)}_{\text{YM}}$}
    \label{tab:bootstrap}
\begin{ruledtabular}
    \begin{tabular}{lcc}
Condition on ${\cal H}^{(2)}_{\text{YM}}$ &  \multicolumn{2}{c}{No. of constraints}  \\  \hline
dimensionless symbols $G$ & 996 & 996 \\ 
no spurious/high-order poles & & \\ 
\hspace{1em} $  
\vev{36}=0$ & 333  & 1189\\
\hspace{1em} $ 
\vev{35}=0$ & 614 & 1546\\
\hspace{1em} $ 
\vev{34}=0$ & 629 & 1916\\
\hspace{1em} $
\vev{45}=0$ & 343& 2116
\\
collinear limit $p_5 || p_6$ 
& & \\
\hspace{1em} $\mathord{-}\mathord{-}\mathord{+}\mathord{+}\mathord{+}\mathord{+}$ & 1785 & 2389\\
\hspace{1em} $\mathord{+}\mathord{-}\mathord{-}\mathord{+}\mathord{+}\mathord{+}$ & 1307 & 2394\\
\hspace{1em} $\mathord{+}\mathord{+}\mathord{-}\mathord{-}\mathord{+}\mathord{+}$ & 1785& 2394 \\
\hspace{1em} 
$\mathord{+}\mathord{+}\mathord{+}\mathord{-}\mathord{-}\mathord{+}$ & 1646 & 2403\\
\hspace{1em} $\mathord{-}\mathord{+}\mathord{+}\mathord{+}\mathord{+}\mathord{-}$ & 1646 & 2403 \\ 
triple collinear limit $p_4||p_5||p_6$ 
&   &
\\ \hspace{1em} 
$\mathord{-}\mathord{-}\mathord{+}\mathord{+}\mathord{+}\mathord{+}$ & 1836 & 2412 \\ 
\hspace{1em}  $\mathord{+}\mathord{+}\mathord{-}\mathord{-}\mathord{+}\mathord{+}$ & 724 & 2412\\  \hline 
{Total} & 2412 & 2412
\end{tabular}
\end{ruledtabular}
\vspace{-2ex}
\end{table}

For six-gluon amplitudes we can also take the triple collinear limit, 
which is defined by
$p_i^{h_i}\to z_i p^h$, $i{=}4,5,6$, with $z_4{+}z_5{+}z_6{=}1$.
In this limit of three external legs becoming collinear, gauge theory amplitudes are expected to factorize into products of splitting functions, and lower point functions, 
\begin{equation}
  \hspace{-0.2 cm}
    \vcenter{\hbox{
 	\begin{tikzpicture}[line width=0.6 ,scale=0.65,line cap=round,every node/.style={font=\footnotesize}]
 			\draw[very thick] (0,0) circle(0.4);
 			\node (tt) at (0,0) {$\mathcal{A}_6$};
 		\begin{scope}[rotate=30]
 			\coordinate (p1) at ($(0.4,0)+1*(0.5,0)$);
 			\coordinate (p2) at ($(0.4,0)$);
  			\node (tt) at ($(p1)+(0.3,0)$) {$p_6$};
 			\draw[very thick] (p1) to (p2);
 		\end{scope}
 		\begin{scope}
 			\coordinate (p1) at ($(0.4,0)+1*(0.5,0)$);
 			\coordinate (p2) at ($(0.4,0)$);
  			\node (tt) at ($(p1)+(0.3,0)$) {$p_5$};
 			\draw[very thick] (p1) to (p2);
 		\end{scope}
 		\begin{scope}[rotate=-30]
 			\coordinate (p1) at ($(0.4,0)+1*(0.5,0)$);
 			\coordinate (p2) at ($(0.4,0)$);
 			 \node (tt) at ($(p1)+(0.3,0)$) {$p_4$};
			\draw[very thick] (p1) to (p2);
 		\end{scope}
 		\begin{scope}[rotate=150]
 			\coordinate (p1) at ($(0.4,0)+1*(0.5,0)$);
 			\coordinate (p2) at ($(0.4,0)$);
 			 \node (tt) at ($(p1)+(0.3,0)$) {$p_1$};
 			\draw[very thick] (p1) to (p2);
 		\end{scope}
 		\begin{scope}[rotate=180]
 			\coordinate (p1) at ($(0.4,0)+1*(0.5,0)$);
 			\coordinate (p2) at ($(0.4,0)$);
 			 \node (tt) at ($(p1)+(0.3,0)$) {$p_2$};
 			\draw[very thick] (p1) to (p2);
 		\end{scope}
 		\begin{scope}[rotate=210]
 			\coordinate (p1) at ($(0.4,0)+1*(0.5,0)$);
 			\coordinate (p2) at ($(0.4,0)$);
 			\node (tt) at ($(p1)+(0.3,0)$) {$p_3$};
 			\draw[very thick] (p1) to (p2);
 		\end{scope}
 	\end{tikzpicture}}}
    \hspace{-3pt}
 	\begin{tikzpicture}[scale=0.7,every node/.style={font=\footnotesize}]
 		\begin{scope}[scale=0.7,every node/.style={font=\footnotesize}]
 				\draw[->,thick] (0,0) to (1,0)
 				node[above] at(0.5,0.1) {$p_4||p_5||p_6$};
 		\end{scope}
 	\end{tikzpicture}\
    \hspace{-9pt}
 	\left(
        \hspace{-4pt}
 		\vcenter{\hbox{
 		\begin{tikzpicture}[scale=0.65,every node/.style={font=\footnotesize}]
 			\begin{scope}
 			\draw[very thick] (0,0) circle(0.4);
 			\node (tt) at (0,0) {$\mathcal{A}_4$};
 		\begin{scope}[rotate=30]
 			\coordinate (p1) at ($(0.4,0)+1*(0.5,0)$);
 			\coordinate (p2) at ($(0.4,0)$);
  			\node (tt) at ($(p1)+(0.3,0)$) {$p$};
 			\draw[very thick] (p1) to (p2);
 		\end{scope}
 		\begin{scope}[rotate=150]
 			\coordinate (p1) at ($(0.4,0)+1*(0.5,0)$);
 			\coordinate (p2) at ($(0.4,0)$);
 			 \node (tt) at ($(p1)+(0.3,0)$) {$p_1$};
 			\draw[very thick] (p1) to (p2);
 		\end{scope}
 		\begin{scope}[rotate=-30]
 			\coordinate (p1) at ($(0.4,0)+1*(0.5,0)$);
 			\coordinate (p2) at ($(0.4,0)$);
 			 \node (tt) at ($(p1)+(0.3,0)$) {$p_3$};
 			\draw[very thick] (p1) to (p2);
 		\end{scope}
 		\begin{scope}[rotate=210]
 			\coordinate (p1) at ($(0.4,0)+1*(0.5,0)$);
 			\coordinate (p2) at ($(0.4,0)$);
 			\node (tt) at ($(p1)+(0.3,0)$) {$p_2$};
 			\draw[very thick] (p1) to (p2);
 		\end{scope}
 		\end{scope}
 	\end{tikzpicture}}}
    \hspace{-3pt}
 	\right)
    \hspace{-3pt}
 	 \times 
    \hspace{-3pt}
 	\vcenter{\hbox{\begin{tikzpicture} [scale=0.75,every node/.style={font=\footnotesize}]		
 		\begin{scope}
 			\coordinate (p1) at (-0.7,0);
 			\coordinate (p2) at ($0.7*({cos(60)},{sin(60)})$);
 			\coordinate (p3) at ($0.7*({cos(-60)},{sin(-60)})$);
 			\coordinate (p4) at ($0.7*(1,0)$);
 			\node (t1) at ($(p1)-(0.2,0)$) {$p$};
 			\node (t2) at ($(p2)+0.2*({cos(60)},{sin(60)})$) {$p_4$};
 			\node (t3) at ($(p3)+0.2*({cos(-60)},{sin(-60)})$) {$p_6$};
 			\node (t4) at ($(p4)+0.2*(1,0)$) {$p_5$};

 			\draw[very thick] (0,0) circle (0.3);
 			\draw[very thick] (p1) to ($(-0.3,0)$);
 			\draw[very thick] (p2) to ($0.3*({cos(60)},{sin(60)})$);
			\draw[very thick] (p3) to ($0.3*({cos(-60)},{sin(-60)})$);
			\draw[very thick] (p4) to ($0.3*({cos(0)},{sin(0)})$);

			\node (t) at (0,0) {$Sp$};
 		\end{scope}
 	\end{tikzpicture}}}
    \, .
 \end{equation}
When applied to the finite hard functions of eq. \eqref{eq:Ampl_IR_subtrac},
this translates to 
\begin{equation}
\begin{aligned}
\mathcal{H}_6^{(2)}/\mathcal{A}_{6}^{(0)} \xrightarrow{ p_4||p_5 || p_6 } &\; \mathcal{H}_4^{(2)}/\mathcal{A}_{4}^{(0)}{+}\mathcal{C}^{h,(1)}_{h_4h_5h_6}\, \mathcal{H}_4^{(1)}/\mathcal{A}_{4}^{(0)}\\
   & {+}\frac12\left(\mathcal{C}^{h,(1)}_{h_4h_5h_6}\right)^2{+}\mathcal{C}^{h,(2)}_{h_4h_5h_6}\label{eq:triplecoH}\,.
\end{aligned}
\end{equation}
$\mathcal{H}_4^{(L)}$ are the  four-point pure YM hard functions of the sector $\mathord{-}\mathord{-}\mathord{+}\mathord{+}$, and $h_4h_5h_6\to h$ can be $\mathord{+}\mathord{+}\mathord{+}\to \mathord{+}$, $\mathord{-}\mathord{+}\mathord{+}\to \mathord{-}$, or $\mathord{+}\mathord{+}\mathord{-}\to \mathord{-}$. The two-loop corrections to the triple splitting functions $\mathcal{C}_{h_4h_5h_6}^{h,(2)}$ are not yet known.
However, we can demand 
that the kinematic dependence of ${\cal H}_6$ factorizes in the limit such that $\mathcal{C}^{h,(2)}_{h_4h_5h_6}$ depends solely on $p_4,p_5,p_6$. 
Moreover, we can impose that $\mathcal{C}^{+,(2)}_{\mathord{+}\mathord{+}\mathord{+}}$ is invariant under swapping $p_4$ and $p_6$. 

Tab.~\ref{tab:bootstrap} summarizes the conditions we impose, with the second column listing the number of constraints each condition places on the initial ansatz. The third column gives the cumulative number of independent constraints obtained by applying the conditions sequentially from top to bottom. At any stage of this process, an inconsistency between the ansatz and the imposed conditions could in principle arise. The fact that all conditions can be satisfied simultaneously therefore provides a non-trivial cross-check of our procedure.

\vspace{-2ex}

\subsection{Including fermions}

So far we have considered the pure YM part of the amplitude. %\jmh{Add the $N_f^2$ term to the equation, and discuss after the eq. why it vanishes at symbol level. Add a footnote mentioning that since it consists of factorized diagrams, it could be relatively easy to compute directly.}
As we are interested in QCD, let us include fermions circulating in the loops. In order to implement the planar limit, we take the number of flavors $N_f$ to be large, while keeping $N_f/N_c$ fixed. As a consequence, QCD hard functions at two loops always have the following expansion,
\begin{equation}
    \mathcal{H}^{(2)}_{\text{QCD}}={\cal H}_{\text{YM}}^{(2)}+\left(\frac{N_f}{N_c}\right)\mathcal{H}^{[1]}+ \left(\frac{N_f}{N_c}\right)^2\mathcal{H}^{[2]}
    \,.
\end{equation}
Compared to the pure-gluon sector, there are fewer physically relevant on-shell diagrams involving fermion loops. As a result, contributions at higher orders in $N_f$ are relatively easier to compute.

For $\mathcal{H}^{[2]}$, the result is IR finite and receives contributions solely from ``kissing box'' on-shell diagrams and their associated integrals~\cite{Carrolo:2026qpu}. What is more, the helicity configuration $\mathord{-}\mathord{-}\mathord{+}\mathord{+}\mathord{+}\mathord{+}$ does not receive any contribution from this cut. Consequently, the maximally transcendental component of $\mathcal{H}^{[2]}$ vanishes \footnote{Note that $\mathcal{H}^{[2]}$ at all transcendentality levels is generated by factorized Feynman diagrams, making it the simplest contribution to compute.}.

For $\mathcal{H}^{[1]}$, we find the following six leading singularities,
\begin{equation}
    S_{i,j}/R_1={-}2+12u_{i,j}-21 u_{i,j}^2+11u_{i,j}^3 \,,
\end{equation}
with $2<i<j\leq6$. Based on this input, we bootstrap the corresponding hard function as described in the previous section. The fermionic bootstrap turns out to be simpler—this sector can be fully determined by enforcing the first three groups of conditions in Table~\ref{tab:bootstrap}; the triple-collinear limit is not required.
\vspace{-2ex}

\subsection{Discussion  of the result}

We obtained the full QCD answer for the maximal-weight symbol of the planar two-loop $\mathord{-}\mathord{-}\mathord{+}\mathord{+}\mathord{+}\mathord{+}$ helicity amplitude. The explicit results are provided in ancillary files. It is worth noting that the functions accompanying the leading singularities $R_{i,j}$ and $S_{i,j}$ with the same indices coincide. This can be understood by the fact that these two sectors can effectively be viewed as components of a six-particle $\mathcal{N}{=}1$ sYM amplitude. 
%This aspect will be explored further in ref. \cite{Carrolo:2026qpu}.
Moreover, since the subtraction, cf. eq. \eqref{eq:sub}, involves two-particle poles only, the hard functions $G_1$ and $G_{i,j}$ 
satisfy the extended Steinmann relations \cite{Caron-Huot:2019bsq,Caron-Huot:2020bkp}. Finally, we find that the symbol alphabet, together with its cyclic permutations, comprises
137 letters. The same set has already been identified in studies of six-point two-loop Wilson loops with Lagrangian insertion \cite{Carrolo:2025pue,Chicherin:2025cua}.

\section{Triple-collinear and double-soft splitting functions}

In the bootstrap procedure, we enforced consistency with triple-collinear limits to fix the result. 
An important byproduct of our result is that we can determine the  triple-collinear splitting functions.
At one loop, the splitting functions were considered in refs. \cite{Catani:2003vu,Sborlini:2014hva,Badger:2015cxa}. 
Thanks to our bootstrap result, we can provide novel information at two loops, on
$\mathcal{C}^{h,(2)}_{h_4h_5h_6}$. 
We find, at maximal transcendental weight level, 
\begin{align}
   &\mathcal{C}^{+,(2)}_{+++}=\mathcal{C}^{(2)}_{\mathcal{N}{=}4\ \text{sYM}}\,,\label{eq:C2ppptop}\\
   &\mathcal{C}^{-,(2)}_{-++}=\mathcal{C}^{+,(2)}_{+++} + \left(r^-_{-++}{+}\frac{N_f}{N_c}s_{-++}^-\right)w \,.\label{eq:C2mpptop}
\end{align}
Here $\mathcal{C}^{(2)}_{\mathcal{N}{=}4\ \text{sYM}}$ is the triple splitting function known from the $\mathcal{N}{=4}$ sYM six-point two-loop remainder function \cite{Bern:2008ap},
$w$ is a weight-four function depending on the two variables $s_{45}/s_{456}$ and $s_{56}/s_{456}$ only, where $s_{ij} = (p_i + p_j)^2$ and $s_{456} = (p_4 + p_5 + p_6)^2$,
and finally
\begin{align}
&r^-_{-++} {=}  
 -2+12u-30u^2+20u^3
    \,,\\
    &s_{-++}^{-}=-2+12u-21u^2+11u^3
\end{align}
where $u$ is given by the triple-collinear limit of $(\vev{35}\vev{46})/(\vev{34}\vev{56})$, depending on the details of the implementation of the limit, see e.g. ref.~\cite{Badger:2015cxa}. 

Another important byproduct of our result are the double-soft splitting functions, which are currently known up to one-loop order \cite{Zhu:2020ftr,Czakon:2022dwk} only. When two external legs $p_5^{h_5}$, $p_6^{h_6}$ are taken to be soft, amplitudes are expected to factorize. At two loops, the six-point hard functions behave under double-soft limits as follows,
\begin{equation}
    \begin{aligned}
\mathcal{H}_6^{(2)}/\mathcal{A}_{6}^{(0)} \xrightarrow{p_5, p_6\to0} &  \mathcal{H}_4^{(2)}/\mathcal{A}_{4}^{(0)}{+}\mathcal{S}^{(1)}_{h_5h_6} \, \mathcal{H}_4^{(1)}/\mathcal{A}_{4}^{(0)} 
\\
   &{+}\frac12\left(\mathcal{S}^{(1)}_{h_5h_6}\right)^2{+}\mathcal{S}^{(2)}_{h_5h_6}\label{eq:dbsoftH} \,.
\end{aligned}
\end{equation}
The maximal-weight part of $\mathcal{S}_{++}^{(2)}$ can be extracted from our bootstrap result.
We find 
\begin{equation}\label{eq:S2ppsoft}
    \mathcal{S}_{++}^{(2)} = \mathcal{S}^{(2)}_{\mathcal{N}{=}4\ \text{sYM}}\,,
\end{equation}
where, remarkably, $\mathcal{S}^{(2)}_{\mathcal{N}{=}4\ \text{sYM}}$ is the double-soft splitting function from the $\mathcal{N}{=}4$ sYM six-point two-loop remainder function.

%\jmh{Finally, it is worth noting that for the full QCD answer, the hard function at $N_f^1$ order goes back exactly to the four-point result, when taking the triple-collinear or double-soft limits, without any correction terms.  Hence, these splitting functions also govern the behavior of full QCD amplitudes.}

% \qinglin{I think it wanted to say when taking any triple-collinear/double-soft limit for $N_f$ term of hard functions, they always go back to $N_f$ term of four-point hard function strictly. Therefore in equations \eqref{eq:triplecoH} or \eqref{eq:dbsoftH}, the last three terms never depend on $N_f$, and splitting function we get from pure YM directly gives the full result for QCD. But this is only true for triple collinear $+++\to +$ or double-soft $++\to0$., while for $-++\to-$, corrections depend on $N_f$. Our long paper Eq.(7.32) make this point correct. I correct some equations according to the long paper in this part.}

The results for the splitting functions of eqs. \eqref{eq:C2ppptop}, \eqref{eq:C2mpptop} and \eqref{eq:S2ppsoft} are recorded in the ancillary files.

\section{Discussion and outlook}

We bootstrapped, for the first time, the symbol of the planar two-loop $\mathord{-}\mathord{-}\mathord{+}\mathord{+}\mathord{+}\mathord{+}$ scattering amplitude in QCD, at maximal transcendental weight. While this is the most complicated part of the amplitude in terms of special functions, we showed that it is also the simplest part in terms of coefficients. The fact that we were able to find an answer that is fully consistent with various physical requirements can be taken as evidence for the correctness of the result, and in particular of the conjecture that the prefactors at maximal weight can be obtained from four-dimensional leading singularities. 
We provide our results in computer-readable form in ancillary files.

Although the total two-loop alphabet comprises $245$ letters, of which $167$ can appear up to weight four, we find that only $137$ of them appear in the final answer.
We take this observation as a hint for further structure, which is yet to be understood. The first steps in this direction have been made in two recent papers that suggest connections of some of the six-particle letters to flag varieties, cf. \cite{Pokraka:2025ali,Bossinger:2025rhf}.

The symbol of the triple collinear splitting function $g \longrightarrow ggg$ that we provide predicts a part of the leading-color two-loop amplitude $gg \to H gg$. 
Similarly, we obtained a new result for the symbol of the planar double soft function, extending previous one-loop results \cite{Zhu:2020ftr,Czakon:2022dwk}. In view of its phenomenological relevance, it would be very interesting to extend these results to the function level, and to include the lower-weight terms.

There are several natural extensions of our work.
\begin{enumerate}
\item Our bootstrap framework extends to other helicity sectors and to processes with external fermions. In upcoming work, we explicitly work out the result for all MHV helicity configurations \cite{Carrolo:2026qpu}. 
The next logical step is to consider next-to-MHV helicity configurations, where we expect additional alphabet letters from refs. \cite{Abreu:2024fei,Henn:2025xrc} to be required, and where certain leading singularities related to three-mass cut configurations may appear.
\item In the present work, we benefited from knowledge of the complete two-loop function space, which had been obtained by traditional means. We anticipate that this requirement can be relaxed 
in future work. One possibility is to instead obtain the information for the symbol bootstrap solely from Landau analysis, see e.g. \cite{Fevola:2023kaw,Hannesdottir:2024hke,He:2024fij,Correia:2025wtb}.
Interestingly, experience from the symbol bootstrap program in sYM suggests that it may be easier to fix the ansatz for higher multiplicity \cite{Drummond:2014ffa}.
Another possibility, which we pursue in upcoming work \cite{Carrolo:2026qpu}, is to combine our approach with ideas from prescriptive unitarity \cite{Bourjaily:2017wjl}, which effectively reduces the number of Feynman integrals that need to be considered.
\item The connection between amplitude coefficients and on-shell diagrams opens up a number of exciting new directions. These include applications to the non-planar case,
%~\footnote{We verified that all maximal-weight coefficients appearing in the full-color two-loop five-gluon amplitudes can be obtained (and simplified) using on-shell diagrams.}, 
as well as to arbitrary particle multiplicities, for which closed-form expressions may be obtained—something that is entirely out of reach for conventional methods.  Another promising avenue is to further investigate the effective supersymmetry relations we have identified, along with Grassmannian representations of on-shell diagrams for the QCD case, extending ideas from ref.~\cite{Arkani-Hamed:2012zlh}.
\item We focused on the symbol, since the latter captures the essential structural information. Repeating this bootstrap at the function level can be done at the level of iterated integrals, using the differential equation representation to generate the necessary expansions (cf. for example ref. \cite{Caron-Huot:2020vlo}).
\item Extending our results below the maximal weight is perhaps the most interesting conceptual point. What is the set of prefactors of transcendental functions needed to describe QCD scattering amplitudes? How much information beyond four dimensions, if any, is needed to describe appropriately defined hard parts of scattering amplitudes? 
These questions are even interesting at the one-loop order, where the rational part is known to be related to subtle ``0/0'' effects, cf. e.g. ref. \cite{Bern:1995db}. 
There are a number of interesting approaches to answer these questions, including loop-level recursion relations 
\cite{Dunbar:2019fcq}, novel integrand representations that highlight the transcendental weight structure of scattering amplitudes \cite{Bourjaily:2021ujs}, and the novel surface integral formulation of Yang-Mills integrands \cite{Arkani-Hamed:2024tzl}.
We hope to revisit these questions in the near future.
\end{enumerate}

\section*{Acknowledgments}
QY acknowledges CERN for its hospitality during his visit, where this work was presented in a seminar, and thanks Samuel Abreu, Pier Francesco Monni, Gherardo Vita, Hantian Zhang and Alexander Zhiboedov for their valuable feedback. 
This work is supported by the European Union (ERC, UNIVERSE PLUS, 101118787). Views and opinions expressed are however those of the authors only and do not necessarily reflect those of the European Union or the European Research Council Executive Agency. Neither the European Union nor the granting authority can be held responsible for them. DC is supported by ANR-24-CE31-7996.  YZ is supported by NSFC through Grant No. 12575078 and 12247103, and thanks Simon Badger and Yu Wu for enlightening discussions.

%\newpage

\bibliographystyle{apsrev4-1}
\bibliography{refs}

@article{Pokraka:2025ali,
    author = "Pokraka, Andrzej and Spradlin, Marcus and Volovich, Anastasia and Weng, He-Chen",
    title = "{Symbol Alphabets in QCD and Flag Cluster Algebras}",
    eprint = "2506.11895",
    archivePrefix = "arXiv",
    primaryClass = "hep-th",
    month = "6",
    year = "2025"
}

@article{Bossinger:2025rhf,
    author = {Bossinger, Lara and Drummond, James and Glew, Ross and G{\"u}rdo{\u{g}}an, {\"O}mer and Wright, Rowan},
    title = "{Singularities of massless scattering and cluster algebras}",
    eprint = "2507.01015",
    archivePrefix = "arXiv",
    primaryClass = "hep-th",
    month = "7",
    year = "2025"
}

@article{Carrolo:2025pue,
    author = "Carr{\^o}lo, S{\'e}rgio and Chicherin, Dmitry and Henn, Johannes and Yang, Qinglin and Zhang, Yang",
    title = "{Hexagonal Wilson loop with Lagrangian insertion at two loops in $ \mathcal{N} $ = 4 super Yang-Mills theory}",
    eprint = "2505.01245",
    archivePrefix = "arXiv",
    primaryClass = "hep-th",
    doi = "10.1007/JHEP07(2025)214",
    journal = "JHEP",
    volume = "07",
    pages = "214",
    year = "2025"
}

@article{Gehrmann:2011xn,
    author = "Gehrmann, Thomas and Henn, Johannes M. and Huber, Tobias",
    title = "{The three-loop form factor in N=4 super Yang-Mills}",
    eprint = "1112.4524",
    archivePrefix = "arXiv",
    primaryClass = "hep-th",
    reportNumber = "HU-EP-11-11-61, NSF-KITP-11-268, ZU-TH-28-11, SI-HEP-2011-19",
    doi = "10.1007/JHEP03(2012)101",
    journal = "JHEP",
    volume = "03",
    pages = "101",
    year = "2012"
}

@article{DeLaurentis:2025dxw,
    author = "De Laurentis, Giuseppe and Ita, Harald and Page, Ben and Sotnikov, Vasily",
    title = "{Compact two-loop QCD corrections for Vjj production in proton collisions}",
    eprint = "2503.10595",
    archivePrefix = "arXiv",
    primaryClass = "hep-ph",
    reportNumber = "PSI-PR-24-29, ZU-TH 14/25",
    doi = "10.1007/JHEP06(2025)093",
    journal = "JHEP",
    volume = "06",
    pages = "093",
    year = "2025"
}

@article{Huss:2025nlt,
    author = {Huss, Alexander and Huston, Joey and Jones, Stephen and Pellen, Mathieu and R{\"o}ntsch, Raoul},
    title = "{Les Houches 2023 -- Physics at TeV Colliders: Report on the Standard Model Precision Wishlist}",
    eprint = "2504.06689",
    archivePrefix = "arXiv",
    primaryClass = "hep-ph",
    reportNumber = "CERN-TH-2025-071, FR-PHENO-2025-005, IPPP/25/19, TIF-UNIMI-2025-9",
    month = "4",
    year = "2025"
}

@article{Henn:2025xrc,
    author = "Henn, Johannes and Matija{\v{s}}i{\'c}, Antonela and Miczajka, Julian and Peraro, Tiziano and Xu, Yingxuan and Zhang, Yang",
    title = "{Complete Function Space for Planar Two-Loop Six-Particle Scattering Amplitudes}",
    eprint = "2501.01847",
    archivePrefix = "arXiv",
    primaryClass = "hep-ph",
    reportNumber = "HU-EP-25/01-RTG, MITP-25-001, MPP-2025-2, USTC-ICTS/PCFT-25-01",
    doi = "10.1103/zhzd-tj9p",
    journal = "Phys. Rev. Lett.",
    volume = "135",
    number = "3",
    pages = "031601",
    year = "2025"
}

@article{Abreu:2024fei,
    author = "Abreu, Samuel and Monni, Pier Francesco and Page, Ben and Usovitsch, Johann",
    title = "{Planar six-point Feynman integrals for four-dimensional gauge theories}",
    eprint = "2412.19884",
    archivePrefix = "arXiv",
    primaryClass = "hep-ph",
    reportNumber = "CERN-TH-2024-221, HU-EP-24/40-RTG",
    doi = "10.1007/JHEP06(2025)112",
    journal = "JHEP",
    volume = "06",
    pages = "112",
    year = "2025"
}

@book{Elvang:2015rqa,
    author = "Elvang, Henriette and Huang, Yu-tin",
    title = "{Scattering Amplitudes in Gauge Theory and Gravity}",
    isbn = "978-1-316-19142-2, 978-1-107-06925-1",
    publisher = "Cambridge University Press",
    month = "4",
    year = "2015"
}

@article{Henn:2021aco,
    author = "Henn, Johannes M. and Bobadilla, William J. Torres",
    title = "{Maximal transcendental weight contribution of scattering amplitudes}",
    eprint = "2112.08900",
    archivePrefix = "arXiv",
    primaryClass = "hep-th",
    reportNumber = "MPP-2021-209",
    doi = "10.1007/JHEP03(2022)174",
    journal = "JHEP",
    volume = "03",
    pages = "174",
    year = "2022"
}

@article{Witten:2003nn,
    author = "Witten, Edward",
    title = "{Perturbative gauge theory as a string theory in twistor space}",
    eprint = "hep-th/0312171",
    archivePrefix = "arXiv",
    doi = "10.1007/s00220-004-1187-3",
    journal = "Commun. Math. Phys.",
    volume = "252",
    pages = "189--258",
    year = "2004"
}

@article{Agarwal:2023suw,
    author = "Agarwal, Bakul and Buccioni, Federico and Devoto, Federica and Gambuti, Giulio and von Manteuffel, Andreas and Tancredi, Lorenzo",
    title = "{Five-parton scattering in QCD at two loops}",
    eprint = "2311.09870",
    archivePrefix = "arXiv",
    primaryClass = "hep-ph",
    reportNumber = "KA-TP-27-2023, MSUHEP-23-030, OUTP-23-12P, P3H-23-088,
  TUM-HEP-1481/23",
    doi = "10.1103/PhysRevD.109.094025",
    journal = "Phys. Rev. D",
    volume = "109",
    number = "9",
    pages = "094025",
    year = "2024"
}

@article{Abreu:2019odu,
    author = "Abreu, S. and Dormans, J. and Febres Cordero, F. and Ita, H. and Page, B. and Sotnikov, V.",
    title = "{Analytic Form of the Planar Two-Loop Five-Parton Scattering Amplitudes in QCD}",
    eprint = "1904.00945",
    archivePrefix = "arXiv",
    primaryClass = "hep-ph",
    reportNumber = "CP3-19-13, FR-PHENO-2019-002, IPhT-19-024",
    doi = "10.1007/JHEP05(2019)084",
    journal = "JHEP",
    volume = "05",
    pages = "084",
    year = "2019"
}

@article{Bourjaily:2021ujs,
    author = "Bourjaily, Jacob L. and Herrmann, Enrico and Langer, Cameron and Patatoukos, Kokkimidis and Trnka, Jaroslav and Zheng, Minshan",
    title = "{Integrands of less-supersymmetric Yang-Mills at one loop}",
    eprint = "2112.06901",
    archivePrefix = "arXiv",
    primaryClass = "hep-th",
    doi = "10.1007/JHEP03(2022)126",
    journal = "JHEP",
    volume = "03",
    pages = "126",
    year = "2022"
}

@article{Bourjaily:2017wjl,
    author = "Bourjaily, Jacob L. and Herrmann, Enrico and Trnka, Jaroslav",
    title = "{Prescriptive Unitarity}",
    eprint = "1704.05460",
    archivePrefix = "arXiv",
    primaryClass = "hep-th",
    reportNumber = "CALT-TH-2017-19",
    doi = "10.1007/JHEP06(2017)059",
    journal = "JHEP",
    volume = "06",
    pages = "059",
    year = "2017"
}

@article{Fevola:2023kaw,
    author = "Fevola, Claudia and Mizera, Sebastian and Telen, Simon",
    title = "{Landau Singularities Revisited: Computational Algebraic Geometry for Feynman Integrals}",
    eprint = "2311.14669",
    archivePrefix = "arXiv",
    primaryClass = "hep-th",
    doi = "10.1103/PhysRevLett.132.101601",
    journal = "Phys. Rev. Lett.",
    volume = "132",
    number = "10",
    pages = "101601",
    year = "2024"
}

@article{Caron-Huot:2020bkp,
    author = {Caron-Huot, Simon and Dixon, Lance J. and Drummond, James M. and Dulat, Falko and Foster, Jack and G\"urdo\u{g}an, \"Omer and von Hippel, Matt and McLeod, Andrew J. and Papathanasiou, Georgios},
    title = "{The Steinmann Cluster Bootstrap for $N$ = 4 Super Yang-Mills Amplitudes}",
    eprint = "2005.06735",
    archivePrefix = "arXiv",
    primaryClass = "hep-th",
    reportNumber = "DESY-20-087",
    doi = "10.22323/1.376.0003",
    journal = "PoS",
    volume = "CORFU2019",
    pages = "003",
    year = "2020"
}

@article{Aybat:2006mz,
    author = "Aybat, S. Mert and Dixon, Lance J. and Sterman, George F.",
    title = "{The Two-loop soft anomalous dimension matrix and resummation at next-to-next-to leading pole}",
    eprint = "hep-ph/0607309",
    archivePrefix = "arXiv",
    reportNumber = "YITP-SB-06-28, SLAC-PUB-11969",
    doi = "10.1103/PhysRevD.74.074004",
    journal = "Phys. Rev. D",
    volume = "74",
    pages = "074004",
    year = "2006"
}

@book{Arkani-Hamed:2012zlh,
    author = "Arkani-Hamed, Nima and Bourjaily, Jacob L. and Cachazo, Freddy and Goncharov, Alexander B. and Postnikov, Alexander and Trnka, Jaroslav",
    title = "{Grassmannian Geometry of Scattering Amplitudes}",
    eprint = "1212.5605",
    archivePrefix = "arXiv",
    primaryClass = "hep-th",
    reportNumber = "PUPT-2435",
    doi = "10.1017/CBO9781316091548",
    isbn = "978-1-107-08658-6, 978-1-316-57296-2",
    publisher = "Cambridge University Press",
    month = "4",
    year = "2016"
}

@article{Badger:2023eqz,
    author = "Badger, Simon and Henn, Johannes and Plefka, Jan Christoph and Zoia, Simone",
    title = "{Scattering Amplitudes in Quantum Field Theory}",
    eprint = "2306.05976",
    archivePrefix = "arXiv",
    primaryClass = "hep-th",
    doi = "10.1007/978-3-031-46987-9",
    journal = "Lect. Notes Phys.",
    volume = "1021",
    pages = "pp.",
    year = "2024"
}

@article{Goncharov:2010jf,
    author = "Goncharov, Alexander B. and Spradlin, Marcus and Vergu, C. and Volovich, Anastasia",
    title = "{Classical Polylogarithms for Amplitudes and Wilson Loops}",
    eprint = "1006.5703",
    archivePrefix = "arXiv",
    primaryClass = "hep-th",
    reportNumber = "BROWN-HET-1602",
    doi = "10.1103/PhysRevLett.105.151605",
    journal = "Phys. Rev. Lett.",
    volume = "105",
    pages = "151605",
    year = "2010"
}

@article{Kotikov:2002ab,
    author = "Kotikov, A. V. and Lipatov, L. N.",
    title = "{DGLAP and BFKL equations in the $N=4$ supersymmetric gauge theory}",
    eprint = "hep-ph/0208220",
    archivePrefix = "arXiv",
    doi = "10.1016/S0550-3213(03)00264-5",
    journal = "Nucl. Phys. B",
    volume = "661",
    pages = "19--61",
    year = "2003",
    note = "[Erratum: Nucl.Phys.B 685, 405--407 (2004)]"
}

@article{Brown:2009qja,
    author = "Brown, Francis C. S.",
    title = "{Multiple zeta values and periods of moduli spaces M 0 ,n ( R )}",
    eprint = "math/0606419",
    archivePrefix = "arXiv",
    journal = "Annales Sci. Ecole Norm. Sup.",
    volume = "42",
    pages = "371",
    year = "2009"
}

@article{Abreu:2019rpt,
    author = "Abreu, Samuel and Dixon, Lance J. and Herrmann, Enrico and Page, Ben and Zeng, Mao",
    title = "{The two-loop five-point amplitude in $ \mathcal{N} $ = 8 supergravity}",
    eprint = "1901.08563",
    archivePrefix = "arXiv",
    primaryClass = "hep-th",
    reportNumber = "CP3-19-06, IPhT-19/003, SLAC-PUB-17377, HU-EP-19/01",
    doi = "10.1007/JHEP03(2019)123",
    journal = "JHEP",
    volume = "03",
    pages = "123",
    year = "2019"
}

@article{Abreu:2018aqd,
    author = "Abreu, Samuel and Dixon, Lance J. and Herrmann, Enrico and Page, Ben and Zeng, Mao",
    title = "{The two-loop five-point amplitude in $\mathcal{N} =4$ super-Yang-Mills theory}",
    eprint = "1812.08941",
    archivePrefix = "arXiv",
    primaryClass = "hep-th",
    reportNumber = "CP3-18-31, IPhT-18/170, SLAC-PUB-17369",
    doi = "10.1103/PhysRevLett.122.121603",
    journal = "Phys. Rev. Lett.",
    volume = "122",
    number = "12",
    pages = "121603",
    year = "2019"
}

@article{Chicherin:2019xeg,
    author = "Chicherin, Dmitry and Gehrmann, Thomas and Henn, Johannes M. and Wasser, Pascal and Zhang, Yang and Zoia, Simone",
    title = "{The two-loop five-particle amplitude in $ \mathcal{N} $ = 8 supergravity}",
    eprint = "1901.05932",
    archivePrefix = "arXiv",
    primaryClass = "hep-th",
    reportNumber = "MPP-2019-5, ZU-TH 04/19, MITP/19-005",
    doi = "10.1007/JHEP03(2019)115",
    journal = "JHEP",
    volume = "03",
    pages = "115",
    year = "2019"
}

@article{Kotikov:2004er,
    author = "Kotikov, A. V. and Lipatov, L. N. and Onishchenko, A. I. and Velizhanin, V. N.",
    title = "{Three loop universal anomalous dimension of the Wilson operators in $N=4$ SUSY Yang-Mills model}",
    eprint = "hep-th/0404092",
    archivePrefix = "arXiv",
    reportNumber = "WSU-HEP-0405",
    doi = "10.1016/j.physletb.2004.05.078",
    journal = "Phys. Lett. B",
    volume = "595",
    pages = "521--529",
    year = "2004",
    note = "[Erratum: Phys.Lett.B 632, 754--756 (2006)]"
}

@article{Heller:2021qkz,
    author = "Heller, Matthias and von Manteuffel, Andreas",
    title = "{MultivariateApart: Generalized partial fractions}",
    eprint = "2101.08283",
    archivePrefix = "arXiv",
    primaryClass = "cs.SC",
    reportNumber = "MITP/21-002, MSUHEP-20-016",
    doi = "10.1016/j.cpc.2021.108174",
    journal = "Comput. Phys. Commun.",
    volume = "271",
    pages = "108174",
    year = "2022"
}

@article{Chicherin:2018old,
    author = "Chicherin, D. and Gehrmann, T. and Henn, J. M. and Wasser, P. and Zhang, Y. and Zoia, S.",
    title = "{All Master Integrals for Three-Jet Production at Next-to-Next-to-Leading Order}",
    eprint = "1812.11160",
    archivePrefix = "arXiv",
    primaryClass = "hep-ph",
    doi = "10.1103/PhysRevLett.123.041603",
    journal = "Phys. Rev. Lett.",
    volume = "123",
    number = "4",
    pages = "041603",
    year = "2019"
}

@article{Arkani-Hamed:2024tzl,
    author = "Arkani-Hamed, Nima and Cao, Qu and Dong, Jin and Figueiredo, Carolina and He, Song",
    title = "{Surface Kinematics and the Canonical Yang-Mills All-Loop Integrand}",
    eprint = "2408.11891",
    archivePrefix = "arXiv",
    primaryClass = "hep-th",
    doi = "10.1103/PhysRevLett.134.171601",
    journal = "Phys. Rev. Lett.",
    volume = "134",
    number = "17",
    pages = "171601",
    year = "2025"
}

@article{Dunbar:2019fcq,
    author = "Dunbar, David C. and Godwin, John H. and Perkins, Warren B. and Strong, Joseph M. W.",
    title = "{Color Dressed Unitarity and Recursion for Yang-Mills Two-Loop All-Plus Amplitudes}",
    eprint = "1911.06547",
    archivePrefix = "arXiv",
    primaryClass = "hep-ph",
    doi = "10.1103/PhysRevD.101.016009",
    journal = "Phys. Rev. D",
    volume = "101",
    number = "1",
    pages = "016009",
    year = "2020"
}

@article{Bern:1994cg,
    author = "Bern, Zvi and Dixon, Lance J. and Dunbar, David C. and Kosower, David A.",
    title = "{Fusing gauge theory tree amplitudes into loop amplitudes}",
    eprint = "hep-ph/9409265",
    archivePrefix = "arXiv",
    reportNumber = "SLAC-PUB-6563, SACLAY-SPH-T-94-95, UCLA-TEP-94-29, SWAT-94-36",
    doi = "10.1016/0550-3213(94)00488-Z",
    journal = "Nucl. Phys. B",
    volume = "435",
    pages = "59--101",
    year = "1995"
}

@article{Zhu:2020ftr,
    author = "Zhu, Yu Jiao",
    title = "{Double soft current at one-loop in QCD}",
    eprint = "2009.08919",
    archivePrefix = "arXiv",
    primaryClass = "hep-ph",
    month = "9",
    year = "2020"
}

@unpublished{DHMZ2021,
  author       = {Nikolaos Dometzgolou and Johannes M. Henn and Julian Miczajka and Simone Zoia},
  title        = {Conformal {Symmetry} of {Leading} {Singularities} of {One-loop} {Scattering} {Amplitudes}},
  year         = {2021},
  note         = {Unpublished.},
}

@article{Drummond:2014ffa,
    author = "Drummond, James M. and Papathanasiou, Georgios and Spradlin, Marcus",
    title = "{A Symbol of Uniqueness: The Cluster Bootstrap for the 3-Loop MHV Heptagon}",
    eprint = "1412.3763",
    archivePrefix = "arXiv",
    primaryClass = "hep-th",
    reportNumber = "CERN-PH-TH-2014-256, LAPTH-232-14",
    doi = "10.1007/JHEP03(2015)072",
    journal = "JHEP",
    volume = "03",
    pages = "072",
    year = "2015"
}

@article{Caron-Huot:2020vlo,
    author = "Caron-Huot, Simon and Chicherin, Dmitry and Henn, Johannes and Zhang, Yang and Zoia, Simone",
    title = "{Multi-Regge Limit of the Two-Loop Five-Point Amplitudes in $\mathcal{N} = 4$ Super Yang-Mills and $\mathcal{N} = 8$ Supergravity}",
    eprint = "2003.03120",
    archivePrefix = "arXiv",
    primaryClass = "hep-th",
    doi = "10.1007/JHEP10(2020)188",
    journal = "JHEP",
    volume = "10",
    pages = "188",
    year = "2020"
}

@article{Carrolo:2026qpu,
    author = "Carr{\^o}lo, S{\'e}rgio and Chicherin, Dmitry and Henn, Johannes and Yang, Qinglin and Zhang, Yang",
    title = "{QCD Scattering Amplitudes and Prescriptive Unitarity}",
    eprint = "2602.02783",
    archivePrefix = "arXiv",
    primaryClass = "hep-th",
    reportNumber = "LAPTH-005/26, MPP-2026-2, USTC-ICTS/PCFT-26-09",
    month = "2",
    year = "2026"
}

@article{Bern:1995db,
    author = "Bern, Z. and Morgan, A. G.",
    title = "{Massive loop amplitudes from unitarity}",
    eprint = "hep-ph/9511336",
    archivePrefix = "arXiv",
    reportNumber = "UCLA-95-TEP-37",
    doi = "10.1016/0550-3213(96)00078-8",
    journal = "Nucl. Phys. B",
    volume = "467",
    pages = "479--509",
    year = "1996"
}

@article{Correia:2025wtb,
    author = "Correia, Miguel and Giroux, Mathieu and Mizera, Sebastian",
    title = "{SOFIA: Singularities of Feynman integrals automatized}",
    eprint = "2503.16601",
    archivePrefix = "arXiv",
    primaryClass = "hep-th",
    doi = "10.1016/j.cpc.2025.109970",
    journal = "Comput. Phys. Commun.",
    volume = "320",
    pages = "109970",
    year = "2026"
}

@article{Hannesdottir:2024hke,
    author = "Hannesdottir, Holmfridur S. and McLeod, Andrew J. and Schwartz, Matthew D. and Vergu, Cristian",
    title = "{Applications of the Landau bootstrap}",
    eprint = "2410.02424",
    archivePrefix = "arXiv",
    primaryClass = "hep-ph",
    doi = "10.1103/PhysRevD.111.085003",
    journal = "Phys. Rev. D",
    volume = "111",
    number = "8",
    pages = "085003",
    year = "2025"
}

@article{He:2024fij,
    author = "He, Song and Jiang, Xuhang and Liu, Jiahao and Yang, Qinglin",
    title = "{Landau-based Schubert analysis}",
    eprint = "2410.11423",
    archivePrefix = "arXiv",
    primaryClass = "hep-th",
    month = "10",
    year = "2024"
}

@article{Czakon:2022dwk,
    author = "Czakon, Micha{\l} and Eschment, Felix and Schellenberger, Tom",
    title = "{Revisiting the double-soft asymptotics of one-loop amplitudes in massless QCD}",
    eprint = "2211.06465",
    archivePrefix = "arXiv",
    primaryClass = "hep-ph",
    doi = "10.1007/JHEP04(2023)065",
    journal = "JHEP",
    volume = "04",
    pages = "065",
    year = "2023"
}

@article{Bern:2008ap,
    author = "Bern, Z. and Dixon, L. J. and Kosower, D. A. and Roiban, R. and Spradlin, M. and Vergu, C. and Volovich, A.",
    title = "{The Two-Loop Six-Gluon MHV Amplitude in Maximally Supersymmetric Yang-Mills Theory}",
    eprint = "0803.1465",
    archivePrefix = "arXiv",
    primaryClass = "hep-th",
    reportNumber = "SLAC-PUB-13150, SACLAY-IPHT-T08-045, UCLA-08-TEP-5, BROWN-HET-1495",
    doi = "10.1103/PhysRevD.78.045007",
    journal = "Phys. Rev. D",
    volume = "78",
    pages = "045007",
    year = "2008"
}

@article{Badger:2015cxa,
    author = "Badger, Simon and Buciuni, Francesco and Peraro, Tiziano",
    title = "{One-loop triple collinear splitting amplitudes in QCD}",
    eprint = "1507.05070",
    archivePrefix = "arXiv",
    primaryClass = "hep-ph",
    reportNumber = "EDINBURGH-2015-12",
    doi = "10.1007/JHEP09(2015)188",
    journal = "JHEP",
    volume = "09",
    pages = "188",
    year = "2015"
}

@article{Sborlini:2014hva,
    author = "Sborlini, German F. R.",
    editor = "Aguilar-Ben{\'\i}tez, M and Fuster, J and Mart{\'\i}-Garc{\'\i}a, S and Santamar{\'\i}a, A",
    title = "{NLO QCD corrections to triple collinear splitting functions}",
    eprint = "1410.1680",
    archivePrefix = "arXiv",
    primaryClass = "hep-ph",
    doi = "10.1016/j.nuclphysbps.2015.09.324",
    journal = "Nucl. Part. Phys. Proc.",
    volume = "273-275",
    pages = "2003--2008",
    year = "2016"
}

@article{Bern:2004cz,
    author = "Bern, Zvi and Dixon, Lance J. and Kosower, David A.",
    title = "{Two-loop g ---{\ensuremath{>}} gg splitting amplitudes in QCD}",
    eprint = "hep-ph/0404293",
    archivePrefix = "arXiv",
    reportNumber = "SLAC-PUB-10414, UCLA-04-TEP-12, SACLAY-SPHT-T04-051, NSF-KITP-04-43",
    doi = "10.1088/1126-6708/2004/08/012",
    journal = "JHEP",
    volume = "08",
    pages = "012",
    year = "2004"
}

@article{Bern:1994zx,
    author = "Bern, Zvi and Dixon, Lance J. and Dunbar, David C. and Kosower, David A.",
    title = "{One loop n point gauge theory amplitudes, unitarity and collinear limits}",
    eprint = "hep-ph/9403226",
    archivePrefix = "arXiv",
    reportNumber = "SLAC-PUB-6415, SACLAY-SPH-T-94-20, UCLA-TEP-94-4, SWAT-94-17",
    doi = "10.1016/0550-3213(94)90179-1",
    journal = "Nucl. Phys. B",
    volume = "425",
    pages = "217--260",
    year = "1994"
}

@article{Kosower:1999rx,
    author = "Kosower, David A. and Uwer, Peter",
    title = "{One loop splitting amplitudes in gauge theory}",
    eprint = "hep-ph/9903515",
    archivePrefix = "arXiv",
    reportNumber = "SACLAY-SPH-T-99-032",
    doi = "10.1016/S0550-3213(99)00583-0",
    journal = "Nucl. Phys. B",
    volume = "563",
    pages = "477--505",
    year = "1999"
}

@article{Catani:2003vu,
    author = "Catani, Stefano and de Florian, Daniel and Rodrigo, German",
    title = "{The Triple collinear limit of one loop QCD amplitudes}",
    eprint = "hep-ph/0312067",
    archivePrefix = "arXiv",
    reportNumber = "CERN-TH-2003-206",
    doi = "10.1016/j.physletb.2004.02.039",
    journal = "Phys. Lett. B",
    volume = "586",
    pages = "323--331",
    year = "2004"
}

@article{Caron-Huot:2019bsq,
    author = "Caron-Huot, Simon and Dixon, Lance J. and Dulat, Falko and Von Hippel, Matt and McLeod, Andrew J. and Papathanasiou, Georgios",
    title = "{The Cosmic Galois Group and Extended Steinmann Relations for Planar $\mathcal{N} = 4$ SYM Amplitudes}",
    eprint = "1906.07116",
    archivePrefix = "arXiv",
    primaryClass = "hep-th",
    reportNumber = "DESY 19-062, DESY-19-062, HU-EP-19/05, SLAC--PUB--17414",
    doi = "10.1007/JHEP09(2019)061",
    journal = "JHEP",
    volume = "09",
    pages = "061",
    year = "2019"
}

@article{Chicherin:2025cua,
    author = "Chicherin, Dmitry and Henn, Johannes and Mazzucchelli, Elia and Trnka, Jaroslav and Yang, Qinglin and Zhang, Shun-Qing",
    title = "{Geometric Landau Analysis and Symbol Bootstrap}",
    eprint = "2508.05443",
    archivePrefix = "arXiv",
    primaryClass = "hep-th",
    reportNumber = "MPP-2025-144, LAPTH-023/25",
    month = "8",
    year = "2025"
}

@article{Catani:1998bh,
    author = "Catani, Stefano",
    title = "{The Singular behavior of QCD amplitudes at two loop order}",
    eprint = "hep-ph/9802439",
    archivePrefix = "arXiv",
    reportNumber = "CERN-TH-98-42, LPTHE-ORSAY-97-57",
    doi = "10.1016/S0370-2693(98)00332-3",
    journal = "Phys. Lett. B",
    volume = "427",
    pages = "161--171",
    year = "1998"
}

\end{document}